\documentstyle[multicol,eqsecnum,aps,pra,epsf]{revtex}
\newcommand{\phm}{\phantom{-}}

\begin{document}

\title{Mixed Heisenberg Chains. II. Thermodynamics.}

\author{Harald Niggemann, Gennadi Uimin\cite{gu}, and Johannes Zittartz}

\address{Institut f\"ur Theoretische Physik, 
Universit\"at zu K\"oln, Z\"ulpicher Str.77, 
D-50937 K\"oln, Germany\\[0.5cm]}

\maketitle

\begin{abstract}
We consider thermodynamic properties, e.g. specific heat, magnetic
susceptibility, of alternating Heisenberg spin chains. Due to a hidden
Ising symmetry these chains can be decomposed into a set of finite
chain fragments. The problem of finding the thermodynamic quantities
is effectively separated into two parts. First we deal with finite
objects, secondly we can incorporate the fragments into a statistical
ensemble. As functions of the coupling constants, the models exhibit
special features in the thermodynamic quantities, e.g. the specific
heat displays double peaks at low enough temperatures. These
features stem from first order quantum phase transitions at zero
temperature, which have been investigated in the first part of this work.
\\[0.8cm]
\end{abstract}

\begin{multicols}{2}
\narrowtext
\section{Introduction}
This work is a continuation of our preceeding paper 
\cite{nzu}, which is devoted to the ground state properties of
alternating Heisenberg spin chains. There we investigated
1$d$ systems described by Hamiltonians (model A)
\begin{eqnarray}
\label{ham_a}
H^{(a)}&=&-{\cal J}_1\sum_{<\!\rho,r\!>}{\bf s}(\rho)\cdot
({\bbox\sigma}(r_1)+{\bbox\sigma}(r_2))\nonumber\\
&&-{\cal J}_0\sum_{<\!r_1,r_2\!>}
{\bbox\sigma}(r_1)\cdot{\bbox\sigma}(r_2),
\end{eqnarray}
or (model B)  
\begin{eqnarray}
H^{(b)}&=&-{\cal J}_1\sum_{<\!\rho,r\!>}({\bf s}(\rho_1)+{\bf s}(\rho_2))\cdot
({\bbox\sigma}(r_1)+{\bbox\sigma}(r_2))\nonumber\\
&&-{\cal J}_0'\sum_{<\!\rho_1,\rho_2\!>}
{\bf s}(\rho_1)\cdot{\bf s}(\rho_2)\nonumber \\
&&-{\cal J}_0\sum_{<\!r_1,r_2\!>}
{\bbox\sigma}(r_1)\cdot{\bbox\sigma}(r_2).\nonumber
\end{eqnarray}
Two kinds of lattice sites, denoted by $\rho$ and $r$, alternate within the
chain. In both models, sites $r_1$ and $r_2$ are occupied by nearest
$\sigma$-spins ($\sigma=1/2$), which can be interpreted as forming a dumbbell
configuration perpendicular to the chain direction. $r$ denotes their common
in-chain coordinate.
In model A, $\rho$ coordinates contain single $s$-spins, whereas
in model B the $\rho$-sites are also occupied by dumbbells of $s$-spins with 
coordinates $\rho_1$ and $\rho_2$. A simple interpretation of model B is
an alternating chain of orthogonal dumbbells.
In this work we concentrate on model A. However, the methods used below
can be reformulated for model B as well.

Two spins, ${\bbox\sigma}(r_1)$ and ${\bbox\sigma}(r_2)$,
are incorporated into the {\it compound spin} 
${\bf S}(r)={\bbox\sigma}(r_1)+{\bbox\sigma}(r_2)$, which is either 0, or 1.
This reveals a hidden Ising symmetry of the original Heisenberg 
model (\ref{ham_a}).
In fact, the ${\cal J}_1$ exchange term in (\ref{ham_a})
does not generate transitions between 
the total spin states 0 and 1 of any compound spin.
Hamiltonian (\ref{ham_a})
can be rewritten in a more suitable form as $H^{(a)}=H_1+H_0^{(a)}$, 
where
\begin{eqnarray}
\label{ham_1}
H_1=-{\cal J}_1\sum_{<\!\rho,r\!>}{\bf s}(\rho)\cdot{\bf S}(r)\\
{\rm and}\quad
H^{(a)}_0=-\frac 12{\cal J}_0\sum_r{\bf S}^2(r).
\label{ham_a0}
\end{eqnarray}
$H^{(a)}_0$ counts the self-energy of a compound spin.

We can use the following evident strategy:
Any configuration of spins is characterized by intrinsic "defects", 
i.e.\ $r$-sites, where the compound spin is zero.
These "defects", which are controlled by the ${\cal J}_0$-terms,
decompose the original chain into an ensemble
of finite chain fragments, which are decoupled from each other.
Their structure can be defined as follows:
A fragment of length $k$ ($k\geq 1$) is an alternating chain of $k+1$
spin-$s$ sites and $k$ spin-1 sites. 
Chain fragments are isolated from each other by a zero spin. 

It is convenient to measure all energies in units of ${\cal J}_1$, the 
latter is supposed to be negative. Thus we set ${\cal J}_1=-1$.

In \cite{nzu} we have observed successive first order transitions 
governed by ${\cal J}_0$ at zero temperature. For model A with $s=1/2$, 
it is a sequence
$\langle 0\rangle \to \langle 1\rangle \to \langle \infty\rangle $, where
a periodicity element $\langle k\rangle$ can be represented
as $(s, 1)^k,s,0$.
E.g., $\langle 0\rangle$ is the periodical alternating structure, where
all $r$-sites are occupied by zero spins.
For spins $s=3/2$ and $s=2$ the phase transition sequence becomes
$\langle 0\rangle \to\langle 1\rangle \to\langle 2\rangle 
\to\langle 3\rangle \to\langle 4\rangle \to\langle \infty\rangle $.
The first two transitions, taking place at ${\cal J}_0^{(0,1)}$ and 
${\cal J}_0^{(1,2)}$, are well isolated from each other, and from 
${\cal J}_0^{(2,3)}$. The latter appears to be very close to the values of
${\cal J}_0^{(3,4)}$ and ${\cal J}_0^{(4,\infty)}$.

A proper method to find out and classify all these
transitions is based on linear programming theory. For our
particular problem it prescribes to compare the reduced energies
$\displaystyle{\frac 1{k+1}}(\epsilon_k - k{\cal J}_0)$
of isolated chain fragments
$(s, 1)^k,s,0$. $\epsilon_k$ is the ground state energy of the 
Hamiltonian $H_1$ (see (\ref{ham_1})) with open boundary conditions. 
It is convenient to introduce the following decomposition of $\epsilon_k$:
\begin{equation}
\label{ener_fit}
\epsilon_k=ke_{\infty}+e_0+e_{\rm int}(k). 
\end{equation}
In (\ref{ener_fit}) $e_{\infty}$ is the energy per element ($s,1$) of the 
perfectly periodic spin structure, $e_0$ is the energy due to the open ends,
and the remaining part, $e_{\rm int}(k)$, 
can be interpreted as the interaction between the chain fragment
ends, which goes to zero at $k\to\infty$. Thus, {\it a succession 
of phase transitions is given by a broken line, which is concave upwards and 
envelops $e_{\rm int}(k)$ from below} \cite{nzu}. Two typical functions
$e_{\rm int}(k)$ are shown in Figs. 1a and 1b. In the former, $e_{\rm int}(k)$ 
is a monotonic function, and thus the system passes through all
intermediate phases
from $\langle 0\rangle$ to $\langle \infty\rangle $, when ${\cal J}_0$
increases from large negative values. In figure 1b, the enveloping function
corresponds to a restricted number of transitions, which 
is typical for mixed Heisenberg chains \cite{nzu}, but with increasing values
of $s$ the minimum becomes very shallow.
Note that if Hamiltonian $H_1$ takes the form of an Ising Hamiltonian, then
$e_{\rm int}(k)\equiv 0$.
In a spin-wave approximation for $s\geq 3/2$, this function 
belongs to the type shown in figure 1a.
\begin{figure}
\epsfxsize=85mm
\epsffile{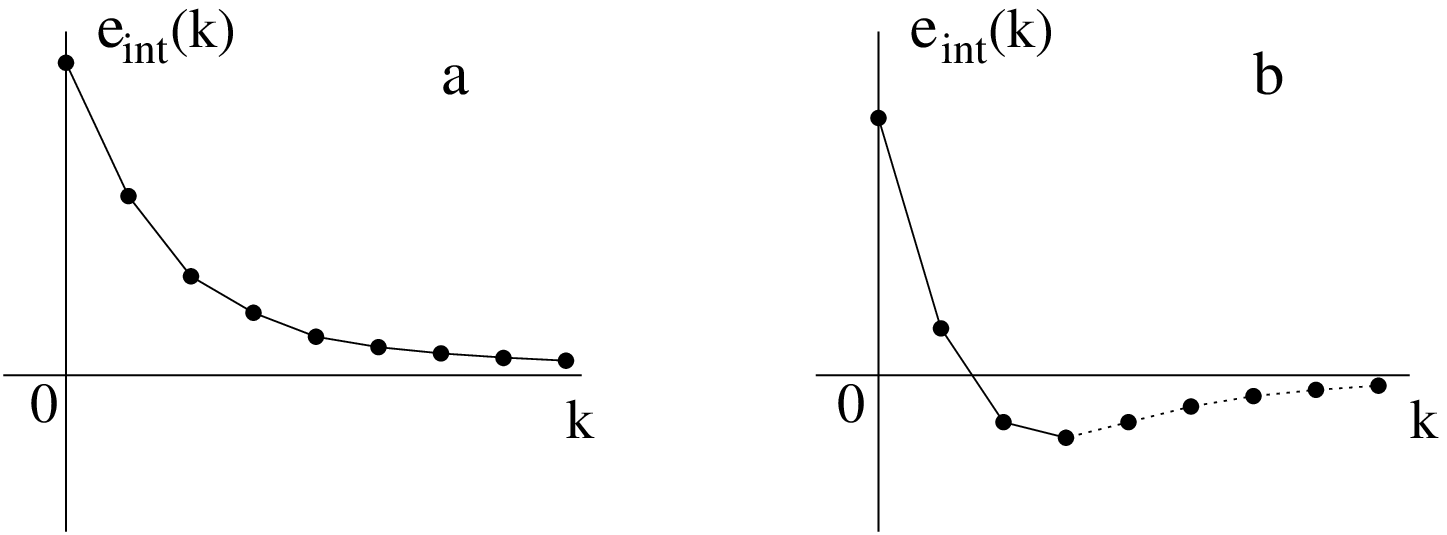}
\hfill
\vspace{-3mm}

\noindent
\caption{Two examples of possible shapes of $e_{\rm int}$ vs $k$:
(a) represents an infinite sequence of transitions,
$\langle 0\rangle \to \langle 1\rangle \to \langle 2\rangle \to \cdots
\to \langle \infty\rangle $. In (b) a finite number of transitions,
$\langle 0\rangle \to \langle 1\rangle \to \langle 2\rangle \to 
\langle 3\rangle \to \langle \infty\rangle $, is realized. The dotted part
of the line does not satisfy the condition "concave upwards".}
\label{fig_1ab}
\end{figure}
Certainly, at non-zero temperature all phase transitions are smeared out.
But thermodynamic quantities may exhibit a crucial dependence on temperature
in vicinities of the critical ${\cal J}_0$ values. We start the next section
with an example which will teach us some peculiarities of the thermodynamics
of these systems.

Besides analytic methods we have used numerical complete
diagonalization of finite chain fragments. By utilizing global $S^z$
conservation we have been able to obtain all energy eigenvalues of the
chain fragments with $k=1,\ldots,6$. The computations have been performed
on an {\em Ultra Enterprise 10000} computer manufactured by
{\em Sun Microsystems}. This is a parallel computer with 40 CPUs and
20 GBytes of shared memory. All programs have been implemented in C++.

\section{Vicinity of the $\langle 0\rangle \to \langle 1\rangle$ transition}
Let us consider the vicinity of the $\langle 0\rangle \to \langle 1\rangle$ 
transition. For $s=1/2$ and zero temperature the system undergoes the 
phase transition at ${\cal J}_0=-2$. At this value the ground state 
is manyfold degenerate: chain fragments of unit length, $k=1$, are embedded 
into the $\langle 0\rangle$ phase. This means that the distributions of
"defects" (zero spins at $r$-sites) is subject to the following constraint:
{\em Two "non-defect" sites (spin-1) cannot be nearest neighbours on the
$r$-sites. They must be separated by at least one "defect".}
The partition function at 
zero temperature is the total number of all valid configurations. 
Let us assume that ${\cal Z}_n$ counts all the configurations, which are
realized between the sites $r=0$ and $r=n$, but the sites $r=0$ and $r=n$
are fixed at zero spin%
\footnote{We label the sites of compound spins by {\it integer numbers} $r$.
For $\rho$, half-integers are reserved.}.
This yields a recurrence relation for the partition function ${\cal Z}$:
\begin{eqnarray}
\label{z1}
{\cal Z}_{n}=2{\cal Z}_{n-1}+{\cal Z}_{n-2}.
\end{eqnarray}
The two terms on the r.h.s. of (\ref{z1}) correspond to the two possibilities
for $r=n\!-\!1$ to be either a "defect" or a "non-defect" site.
If it is a "defect" site (first term), we count all configurations between 
the "defect" sites, $r=0$ and $r=n\!-\!1$. 
The factor 2 is due to the additional spin-1/2 at site
$\rho=n-1/2$. This spin is enclosed by "defects", so it is {\em free} or
paramagnetic.
The second term on the r.h.s. of (\ref{z1}) corresponds to the case, where
$r=n-1$ is a "non-defect" site. Since "non-defects" are not allowed to be
nearest
neighbours, $r=n-2$ must contain a "defect". Thus
the configuration count of the remaining part is ${\cal Z}_{n-2}$.
Evi\-dent\-ly, the boundary conditions for
(\ref{z1}) must be chosen as ${\cal Z}_{0}=1$ and ${\cal Z}_{1}=2$.

Now we define an additional quantity ${\cal D}_n$, which counts the number of
"non-defects" between the spin-singlets on sites $r=0$ and $r=n$, summed over
all allowed configurations. It satisfies the equation
\begin{eqnarray}
\label{d1}
{\cal D}_{n}=2{\cal D}_{n-1}+{\cal D}_{n-2}+{\cal Z}_{n-2}.
\end{eqnarray}
The first two terms on the r.h.s.\ of (\ref{d1}) are
similar to those on the r.h.s.\ of Eq.\ (\ref{z1}).
The third term counts how many times the "non-defect" at $r=n-1$ appears in 
all valid
configurations. For ${\cal D}$, the boundary conditions are
${\cal D}_{0}=0$ and ${\cal D}_{1}=0$.

Let us discuss, how Eqs.\ (\ref{z1})-(\ref{d1}) have to be modified in case of
non-zero temperatures.
If all terms in ${\cal Z}_n$ are preceded by their statistical weights,
we can use this quantity to calculate the partition function of a chain of
length $n$ with open boundary conditions.
Instead of the set $\{\epsilon_k\}$ (cf. (\ref{ener_fit})), one should 
make use of the set of free energies $\{\phi_k\}$. Since we consider the
vicinity of the $\langle 0\rangle \to \langle 1\rangle$ transition
(i.e.\ ${\cal J}_0 \simeq -2$), only $\phi_0$ and $\phi_1$ enter the
calculation. All other configurations are suppressed at low temperatures
($T\ll 1$).
Introducing Boltzmann factors $w_k=\exp-(\phi_k-k{\cal J}_0)/T$, and 
setting $\phi_0=-T\ln 2$ and $\phi_1= -2$, we arrive at the following 
modification of Eqs.\ (\ref{z1})-(\ref{d1}):
\begin{eqnarray}
\label{z2}
{\cal Z}_{n}=2{\cal Z}_{n-1}+w{\cal Z}_{n-2}, \quad w=w_1,
\end{eqnarray}
\begin{eqnarray}
\label{d2}
{\cal D}_{n}=2{\cal D}_{n-1}+w({\cal D}_{n-2}+{\cal Z}_{n-2}),
\end{eqnarray}
The solution of (\ref{z2}) is
$${\cal Z}_{n}=c_+\lambda_+^n+c_-\lambda_-^n, 
\quad \lambda_{\pm}=1\pm \sqrt{1+w}.$$
For solving (\ref{d2}), we try the ansatz
$${\cal D}_{n}=(a_++nb_+)\lambda_+^n+(a_-+nb_-)\lambda_-^n.$$
For the concentration of spin-1 sites in long chains we obtain
\begin{eqnarray}
\label{conc}
x=\lim_{n\to\infty}\frac{{\cal D}_n}{n{\cal Z}_n}=\frac{b_+}{c_+}.
\end{eqnarray}
The relationship between $b_+$ and $c_+$, which enter the 
leading terms of ${\cal D}_{n}$ and ${\cal Z}_{n}$ for $n\gg 1$,
can be directly derived from Eq.\ (\ref{d2}):
$$\frac{b_+}{c_+}=\frac{w}{\lambda_+^2+w}=\frac w{2(1+w+\sqrt{1+w})}.$$ 

The physical quantities, such as the specific heat and the entropy, can 
straightforwardly be calculated as derivatives of the free energy. 
In the thermodynamic limit, i.e.\ $n\to\infty$, the free energy per
compound spin is simply 
\[
-T\ln \lambda_+ = -\frac{2+{\cal J}_0}{2}-\frac{T}{\sqrt{w}},
\]
if $\langle 1\rangle$ is favorable ($w\gg 1$).
Otherwise, for $w\ll 1$ we arrive at the expression
\[
-T\ln 2 -\frac{Tw}{4}.
\]

The magnetic susceptibility reflects the groundstate investigations of our
former paper. In fact,
as far as chain fragments of length $k=1$ are in the spin-singlet state,
only "isolated" spin-$1/2$ sites, associated with $k=0$, contribute a
Curie-like susceptibility. It can be expressed as a contribution of individual
"isolated" spins, 
$\displaystyle{\frac {\mu_B^2}{4T}}$, multiplied by their 
concentration
$$1\!-\!2b_+/c_+=(1+\sqrt{1+w})/(1+w+\sqrt{1+w}).$$ 
This exhibits "half-a-gap" behavior, $\propto 1/\sqrt{w}$, 
when $w\gg 1$, {i.e.}, ${\cal J}_0>-2$. This, and the analogous $w$-dependence
of the specific heat can be interpreted in terms of 
an equilibrium chemical reaction, in which any spin singlet (1/2, 1, 1/2)
may transform into {\it two} paramagnetic spins 1/2 and a "defect".

In the vicinity of the $\langle 0\rangle \to \langle 1\rangle$ transition,
the results of this section are not only valid for $s=1/2$, but also for
other spins on the $\rho$-sites.
Evident changes are:
\begin{enumerate}
\item
$\phi_1=-(2s\!+\!1)$, and ${\cal J}_0$ varying around $-(2s\!+\!1)$.
\item
$\phi_0=-T\ln (2s\!+\!1)$, resulting in
$$\lambda_{\pm}=(s\!+\!1/2)\pm \sqrt{(s\!+\!1/2)^2+w},$$ 
and $$\frac{b_+}{c_+}=\frac{w/2}{(s+1/2)^2+w+(s+1/2)\sqrt{(s+1/2)^2+w}}.$$ 
\item
A group of spins ($s, 1, s$), whose total spin value at low 
temperatures is practically $(2s\!-\!1)$, is paramagnetic too, 
as well as an "isolated" spin $s$.
\end{enumerate}
The Curie-like susceptibility is straightforwardly calculated as
$$ \frac {\mu_B^2}{3T}(\,2s(2s\!-\!1)b_+/c_++s(s\!+\!1)(1-2b_+/c_+)\,).$$
\section{General consideration}
For $s\!=\!1/2$, 
quantum fluctuations are efficient enough to ``isolate'' the
$\langle 1\rangle\! \leftrightarrow \!\langle \infty\rangle$ transition from 
$\langle 0\rangle \! \leftrightarrow \! \langle 1\rangle$.
In Ref.~\cite{nzu} these zero-temperature transitions were estimated as
${\cal J}_0^{(1,\infty)}=-0.910$ 
and ${\cal J}_0^{(0,1)}=-2$, respectively.
Thus, at low temperatures we can investigate the regions around 
${\cal J}_0^{(1,\infty)}$ and ${\cal J}_0^{(0,1)}$ separately.
In Section~II the subject of interest is the
competition of ``defect'' and ``non-defect'' sites, provided two 
``non-defects'' cannot be nearest neighbours, if ${\cal J}_0$ is 
around ${\cal J}_0^{(0,1)}$. On the other hand,
two ``defects'' cannot be nearest neighbours in the second critical range
around ${\cal J}_0^{(1,\infty)}$ at low temperatures.
Instead of dealing with ``defect'' and ``non-defect'' objects, let us use
the convention of Ref.~\cite{nzu}. For convenience, 
chain fragments have been defined as follows: A chain fragment of length $k$ 
formally includes a spin-0 site from its right, so it can be represented as
(1/2,1)$^k$(1/2,0). Conventionally, the nearest spin from the left of any
chain fragment is also 0, but this spin is incorporated into the fragment
which lies on the l.h.s.\ of this site.
In this classification, a chain fragment of zero length is (1/2,0).

Now we can reformulate one of the statements mentioned above.
In the second ``critical'' range around ${\cal J}_0^{(1,\infty)}$ 
zero-length chain fragments are incompetitive
and may be neglected at low temperatures. However, we shall see
from the specific heat calculation 
that at intermediate temperatures, e.g. $T=0.3$, and intermediate
${\cal J}_0$ between ${\cal J}_0^{(0,1)}$ and ${\cal J}_0^{(1,\infty)}$,
not only chain fragments of unit length dictate thermodynamic properties,
but longer chain fragments and zero-length fragments also contribute 
essentially. Therefore we have to take into account chain fragments of any
length, and no special restrictions on the values of ${\cal J}_0$ and
temperature will be imposed.

The free energy of a chain fragment of length $k$ can be written as a
generalization of Eq.\ (\ref{ener_fit}):
\begin{equation}
\label{fr_ener_fit}
\phi_k=kf_{\infty}+f_0+f_{\rm int}(k), 
\end{equation}
where $f_{\infty}, f_0$ and $f_{\rm int}$ are temperature-dependent.
In this section, the configurational part of the free energy will be
determined by making use of recursive relations similar to
(\ref{z2})-(\ref{d2}).

It is convenient to take the global configuration
$\langle \infty\rangle$ as the ``vacuum'' state. A zero spin within this
background is called a ``hole''.
If we ignore the interaction term in (\ref{fr_ener_fit}),
i.e.\ $f_{\rm int}(k)$, then any ``hole'' costs the free energy
$({\cal J}_0-f_{\infty}(T))+f_0(T)$. In fact, the free energy of a very long
chain, $nf_{\infty}(T)+f_0(T)$, becomes $(n-1)f_{\infty}(T)+2f_0(T)+{\cal J}_0$
if a ``hole'' is inserted. For two ``holes'' we obtain
$(n-2)f_{\infty}(T)+3f_0(T)+2{\cal J}_0$, and so on.
However, this consideration is no more valid if two ``holes''
occupy nearest neighbour sites on the $r$-sublattice. Such two ``holes'' give 
rise to a zero length chain fragment. In this case we obtain
$(n-2)f_{\infty}(T)+2f_0(T)+2{\cal J}_0$, which can be subdivided into
$(n-1)f_{\infty}(T)+2f_0(T)+{\cal J}_0$ and $-f_{\infty}(T){\cal J}_0$.
The latter should be interpreted as the free energy of the 
zero length chain fragment. 
Only the ``holes'' which are separated by a chain fragment of non-zero length
and which do not have other ``holes'' between them, interact via 
$f_{\rm int}(k;T)$, where $k\geq 1$ is the chain fragment length, 
or the number of spin-1 sites between these 
{\it nearest} holes. We denote the statistical weight of a chain fragment 
of length $k$ by $w_k$. Then starting with 
$$w_0=\exp \left[-({\cal J}_0-f_{\infty})/T\right]\quad {\rm and}$$
$$w_1=\exp \left[-({\cal J}_0-f_{\infty}+f_0+f_{\rm int}(1))/T\right],$$
we obtain
\begin{eqnarray}
\label{boltz}
w_k=w_1 \exp\left[(f_{\rm int}(1)-f_{\rm int}(k))/T\right], \quad k\!\geq\!1
\end{eqnarray}
for longer chain fragments.

Recurrence relations for the partition function are
evident generalizations of (\ref{z2}):
\begin{eqnarray}
\label{z3}
{\cal Z}_{n+1}=2w_0{\cal Z}_n+w_1{\cal Z}_{n-1}+w_2{\cal Z}_{n-2}
\nonumber\\+\cdots+w_k{\cal Z}_{n-k}
+\cdots+w_n{\cal Z}_{0}, \quad n\geq 1.
\end{eqnarray}
Each ${\cal Z}_k$ counts all possible spin configurations with corresponding
statistical weights between $r\!=\!0$ and $r\!=\!k$, while fixing the boundary
compound spins at $r\!=\!0$ and $r\!=\!k$ at zero.
The prefactor 2 of $w_0{\cal Z}_n$ on the r.h.s.\ of Eq.~(\ref{z3}) is due
to the spin-1/2 at site $\rho=n+1/2$.
The boundary conditions for the set of 
partition functions are ${\cal Z}_0=1$ and ${\cal Z}_1=2w_0$. 
The latter reflects the existence of a free spin-1/2 between two spin-0 
sites.

The lower index in ${\cal D}_{k}^{(m)}$ has the same meaning
as in ${\cal Z}_k$, whereas the upper index is related to the chain 
fragment length. ${\cal D}_{k}^{(m)}$ measures how often chain
fragments of length $m$ occur between the sites $r\!=\!0$ and $r\!=\!k$.
Of course, any spin configuration in ${\cal D}_{k}^{(m)}$ picks up 
a corresponding statistical weight.
The recurrence relations for shortest chain fragments, $k\!=\!0$ and 1, have a
structure which is similar to that of Eq.~(\ref{z3}):
\begin{eqnarray}
\label{d3_0}
{\cal D}_{n+1}^{(0)}=2w_0{\cal Z}_{n}+2w_0{\cal D}_{n}^{(0)}+
w_1{\cal D}_{n-1}^{(0)}
\nonumber\\
+\cdots+w_k{\cal D}_{n-k}^{(0)}
+\cdots+w_n{\cal D}_{0}^{(0)}, \quad n\geq 1.
\end{eqnarray}
\begin{eqnarray}
\label{d3_1}
{\cal D}_{n+1}^{(1)}=w_1{\cal Z}_{n-1}+2w_0{\cal D}_{n}^{(1)}+
w_1{\cal D}_{n-1}^{(1)}
\nonumber\\
+\cdots+w_k{\cal D}_{n-k}^{(1)}
+\cdots+w_n{\cal D}_{0}^{(1)}, \quad n\geq 1.
\end{eqnarray}
The first terms on the r.h.s.\ of (\ref{d3_0}) and (\ref{d3_1})
are the contributions of the spin-0 and spin-1
sites at $r\!=\!n$, respectively.
The boundary condition, which should be imposed on ${\cal D}^{(0)}$, reads
${\cal D}_0^{(0)}=0$. For ${\cal D}^{(1)}$, we can set ${\cal D}_0^{(1)}=0$
and ${\cal D}_1^{(1)}=0$. 

Generalization of the recurrence relations and boundary conditions
to arbitrary chain fragment lengths $m$ is also evident:
\begin{eqnarray}
\label{d4}
{\cal D}_{n+1}^{(m)}=w_m{\cal Z}_{n-m}+2w_0{\cal D}_{n}^{(m)}+
w_1{\cal D}_{n-1}^{(m)}\nonumber\\+
\cdots+w_k{\cal D}_{n-k}^{(m)}+\cdots
+w_n{\cal D}_{0}^{(m)}, \quad n\geq m,\\
{\cal D}_n^{(m)}=0 \quad {\rm for} \quad n\leq m.\nonumber
\end{eqnarray}

We define the concentration $x_m$ of chain fragments of length $m$
as the ratio of the expectation value of their total number
$N_m={\cal D}_n^{(m)}/{\cal Z}_n$ to the total number of $r$-sites, $n$:
\begin{eqnarray}
\label{conc1}
x_m=\lim_{n\to\infty}\frac{{\cal D}_n^{(m)}}{n{\cal Z}_n},
\end{eqnarray}
similar to equation (\ref{conc}).
These concentrations must satisfy the conservation law
\begin{eqnarray}
\label{sr}
1=\sum_{k\geq 0}(k+1)x_k,
\end{eqnarray}
which states that the total number of compound spins, zeros and ones,
is equal to the total number of $r$-sublattice sites.

The set of equations (\ref{z3})-(\ref{d4}) allows us to perform a
straightforward numerical analysis. However, let us try
an analytical approach by assuming that 
$|f_{\rm int}(k)-f_{\rm int}(1)|\ll T$ for all $k\geq 1$.
Practically, this means that 
\begin{equation}
T\gg |f_{\rm int}(1)|,
\label{inequ}
\end{equation}
which is well satisfied at $T\!>\!0.15$, as we shall see in section IV.
In this approximation one can set $w_k=w$ for all $k\geq 1$.
By subtracting ${\cal Z}_n$ from ${\cal Z}_{n+1}$, we rewrite 
Eqs.~(\ref{z3}) in a simple form ($n\geq 1$)
\begin{eqnarray}
\label{z4}
{\cal Z}_{n+1}-(1+2w_0){\cal Z}_n+(2w_0-w_1){\cal Z}_{n-1}
=0.
\end{eqnarray}
Analogous equations for ${\cal D}_k^{(0)}$ and ${\cal D}_k^{(1)}$ 
are similar, but have
different r.h.s.\ and boundary conditions:
\begin{eqnarray}
\label{d5_0}
{\cal D}_{n+1}^{(0)}-(1+2w_0){\cal D}_{n}^{(0)}+(2w_0-w_1){\cal D}_{n-1}^{(0)}
\nonumber\\
=2w_0({\cal Z}_{n}-{\cal Z}_{n-1}), \quad n\geq 1
\\
{\cal D}_0^{(0)}=0,\;{\cal D}_1^{(0)}=2w_0.
\nonumber
\end{eqnarray}
\begin{eqnarray}
\label{d5_1}
{\cal D}_{n+1}^{(1)}-(1+2w_0){\cal D}_{n}^{(1)}+(2w_0-w_1){\cal D}_{n-1}^{(1)}
\nonumber\\
=w_1({\cal Z}_{n-1}-{\cal Z}_{n-2}), \quad n\geq 2
\\
{\cal D}_0^{(1)}={\cal D}_1^{(1)}=0,\;{\cal D}_2^{(1)}=w_1.
\nonumber
\end{eqnarray}
As in the case of Eqs.~(\ref{z2})-(\ref{d2}), we look for solutions
${\cal Z}_n$, ${\cal D}_n^{(0)}$ and ${\cal D}_n^{(1)}$ of the form 
\begin{eqnarray}
c_+\lambda_+^n+c_-\lambda_-^n, \quad
(a_++b_+^{(0)}n)\lambda_+^n+(a_-+b_-^{(0)}n)\lambda_-^n,\nonumber\\ 
{\rm and} \quad 
(a_++b_+^{(1)}n)\lambda_+^n+(a_-+b_-^{(1)}n)\lambda_-^n,\nonumber
\end{eqnarray}
respectively. By inserting this ansatz we obtain
$$\lambda_+=w_0+1/2+\sqrt{2w_0^2+w_1+1/4},$$
and the concentrations of chain fragments of length $k=1$ and 0:
\begin{eqnarray}
\label{b_1}
x_1=\frac{b_+^{(1)}}{c_+}=
\frac {w_1(\lambda_+-1)}{\lambda_+(\lambda_+^2+w_1-2w_0)},
\end{eqnarray}
\begin{eqnarray}
\label{b_0}
x_0=\frac{b_+^{(1)}}{c_+}=\frac {2w_0\lambda_+}{w_1}\,x_1.
\end{eqnarray}

Generalization of Eq.~(\ref{b_1}) for $m\!>\!1$ is straightforward: The
analogue of (\ref{d5_1}) now reads
\begin{eqnarray}
\label{d6}
{\cal D}_{n+1}^{(m)}-(1+2w_0){\cal D}_{n}^{(m)}+(2w_0-w_1){\cal D}_{n-1}^{(m)}
\nonumber\\
=w_1({\cal Z}_{n-m}-{\cal Z}_{n-m-1}), \quad n\geq m\!+\!1,\end{eqnarray}
while the boundary conditions can be written as 
\begin{eqnarray}
{\cal D}_0^{(m)}\!=\dots={\cal D}_m^{(m)}\!=0,
\;{\cal D}_{m+1}^{(m)}\!=w_1.  \nonumber
\end{eqnarray}
The abovementioned asymptotic behaviour of ${\cal Z}_n$ and ${\cal D}_n^{(m)}$
allows us to determine the concentration $x_m$ as
\begin{eqnarray}
\label{b_m}
x_m=\frac{b_+^{(m)}}{c_+}=\frac 1{(\lambda_+)^{m-1}}\,x_1.
\end{eqnarray}
It is not difficult to check the validity of the sum rule (\ref{sr}) with
equations (\ref{b_0}), (\ref{b_1}) and (\ref{b_m}).

Just comparing two sets of equations, (\ref{d5_1}) and (\ref{d6}),
one can conclude that the knowledge of all ${\cal D}_k^{(1)}$
allows us to calculate any expression of higher rank, e.g. 
\begin{eqnarray}
\label{d_1_shift}
{\cal D}_{n+m-1}^{(m)}={\cal D}_n^{(1)}.
\end{eqnarray}

At lower temperatures it is necessary to bring in more Boltzmann factors.
The simplest extension of the temperature range
\begin{equation}
T\gg |f_{\rm int}(2)|
\label{inequ2}
\end{equation}
does not impose restrictions on $|f_{\rm int}(1)|$ any more, the latter
is not necessarily much smaller than the tem\-pe\-ra\-ture.
This leads us to the introduction of three Boltzmann factors, 
$w_0$, $w_1$, and
$w_k=w_2\approx w_1\exp[f_{\rm int}(1)/T]$, 
if $k\!\geq\!2$ (see definition (\ref{boltz})). 
Numerical results of the next section will show
that such a description is a good approximation for $T\!>\!0.04$.
This is a systematic way to extend our approach to lower temperatures.
By using four Boltzmann factors, the system can be described down to
$T\approx 0.01$.

In Appendix A we outline how the method developed above works in the
general situation, where temperature is restricted from below:
$T\gg |f_{\rm int}(k)|$. In this case we deal with a set of statistical 
weights 
\begin{equation}
\label{set_w}
\{w_0, w_1, w_2,\dots,w_{k-1},w_n=w_{k} (n\geq k)\}.
\end{equation}
\section{Numerical computations}
\subsection{Chain fragments: Zero-field results}
By using complete matrix diagonalization we have computed all energy
eigenvalues of finite chain fragments up to $k=6$. From these eigenvalues,
the exact values of the free energy for various temperatures have been
calculated. We decompose the free energy into an affine contribution
$k f_\infty + f_0$ and an "interaction" contribution $f_{\rm{int}}(k)$.
This decomposition is defined by
$ \displaystyle{\lim_{k\to\infty} f_{\rm{int}}(k) = 0.}$ 
\begin{table}
\[ \begin{array}{|c|l|l|l|}
\hline
k & \multicolumn{1}{|c|}{T=0.02} &
    \multicolumn{1}{|c|}{T=0.3}  &
    \multicolumn{1}{|c|}{T=1.0}  \\
\hline
\hline
0 & \phm 0.454050 & \phm 0.245974          & 0.027743 \\
1 & -0.074381     & \phm 0.002104          & 0.000148 \\
2 & -0.012040     & -9.540148\cdot 10^{-5} & 8.060468\cdot 10^{-7}  \\
3 & -0.000113     & -5.336452\cdot 10^{-6} & 4.404270\cdot 10^{-9}  \\
4 & \phm 0.000476 & -8.747215\cdot 10^{-8} & 2.376055\cdot 10^{-11} \\
5 & \phm 0.0      & \phm 0.0               & 0.0 \\
6 & \phm 0.0      & \phm 0.0               & 0.0 \\
\hline
\end{array} \]
\caption{The "interaction" contribution $f_{\rm{int}}(k)$ of the free energy
for $T=0.02$, $T=0.3$, and $T=1.0$.}
\label{tab_free}
\end{table}

\begin{figure}
\epsfxsize=85mm
\epsffile{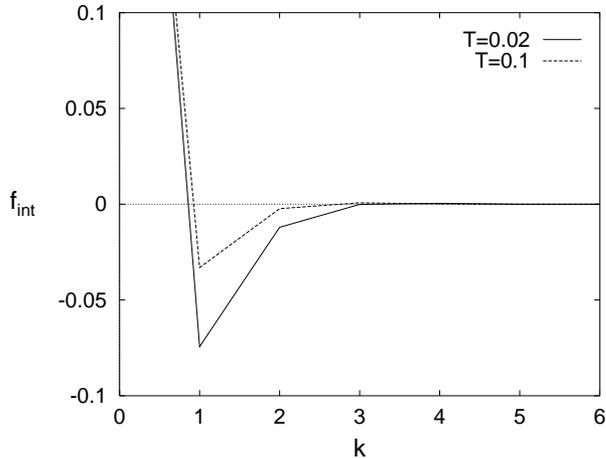}
\vspace{3mm}

\caption{Computed free energy vs $k$ after subtracting the affine contribution
$k f_\infty + f_0$ for low temperatures $T=0.02$ (full) and $T=0.1$ (dashed)}
\label{fig_freelow}
\end{figure}

\begin{figure}
\epsfxsize=85mm
\epsffile{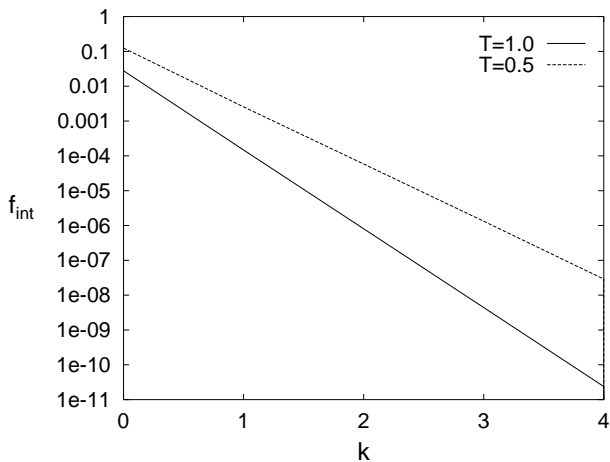}
\vspace{3mm}

\caption{"Interaction part" of free energy as a function of $k$ for high
temperatures $T=1.0$ (full) and $T=0.5$ (dashed)}
\label{fig_freehigh}
\end{figure}

\begin{figure}
\epsfxsize=85mm
\epsffile{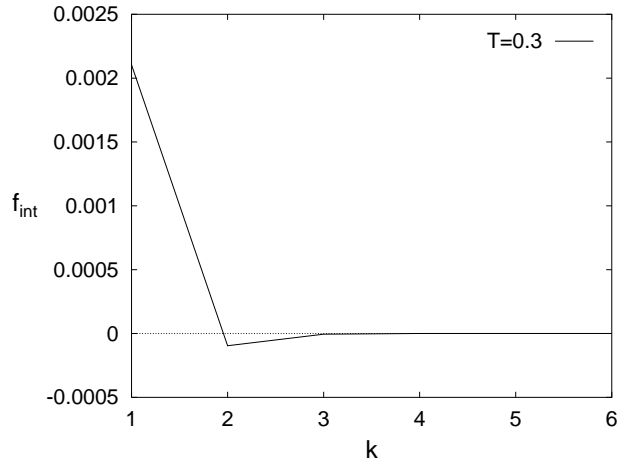}
\vspace{3mm}

\caption{"Interaction part" of free energy as a function of $k$ at the
crossover temperature $T_c\simeq 0.3$}
\label{fig_freeinter}
\end{figure}

Table \ref{tab_free} shows computed values of $f_{\rm{int}}(k)$ for
$T=0.02$, $T=0.3$, and $T=1.0$. Free energy values for $k=5$ and $k=6$ have
been used to determine $f_0$ and $f_\infty$, so
\[ f_{\rm{int}}(5)=f_{\rm{int}}(6)\equiv 0 \]
for all temperatures. The accuracy of this approximation can be estimated
from figures \ref{fig_freelow}, \ref{fig_freehigh}, and \ref{fig_freeinter},
where $f_{\rm{int}}(k)$ decays quickly and smoothly to zero.

For low temperatures ($T<0.3$), the $k$-dependence of the free energy is
qualitatively similar to the ground state energy. This is illustrated in
figure \ref{fig_freelow}. 
As in the ground state energy, there
is a clear minimum at $k=1$, which corresponds to the phase transition
sequence $\langle 0\rangle \to \langle 1\rangle \to \langle \infty\rangle$
at $T=0$.

For $T>0.3$, temperature has smeared out the ground state structure
completely. Deviations from the affine contribution
$k f_\infty + f_0$ decay exponentially fast as a function of $k$.
This is shown in figure \ref{fig_freehigh} for $T=0.5$ and $T=1.0$.

In the crossover region $T\simeq 0.3$ we observe an interesting phenomenon:
As shown in figure \ref{fig_freeinter}, the
minimum of the free energy has moved from $k=1$ to $k=2$. This is due to
the spin-degeneracy of the chain fragment. The total spin of a chain fragment 
of length $k$ is $s_p=(k-1)/2$ for low temperatures, so the
degeneracy is $2s_p + 1 = k$. This yields
\[
f_{\rm{deg}}=-T\,\ln k
\]
as an additional contribution to the free energy. If we subtract this
additional contribution from $f_{\rm{int}}$, the minimum at $k=1$ is
restored. This contribution is {\em only} relevant at intermediate
temperatures: At low temperatures it is suppressed by the prefactor
$T$, at high temperatures the $\ln k$ contribution to the spin entropy 
is no more dominant. Indeed, the magnetic behaviour of short chain fragments
cannot be described by a single paramagnetic spin $s_p$. In other words,
at high temperatures the spin--spin correlation length becomes smaller
than $k$.

\subsection{Chain fragments: Magnetic susceptibility}
By including small magnetic fields into the Hamiltonian, we have computed
the zero-field susceptibility for $k=1,\ldots,5$ and various temperatures.

\begin{figure}
\epsfxsize=85mm
\epsffile{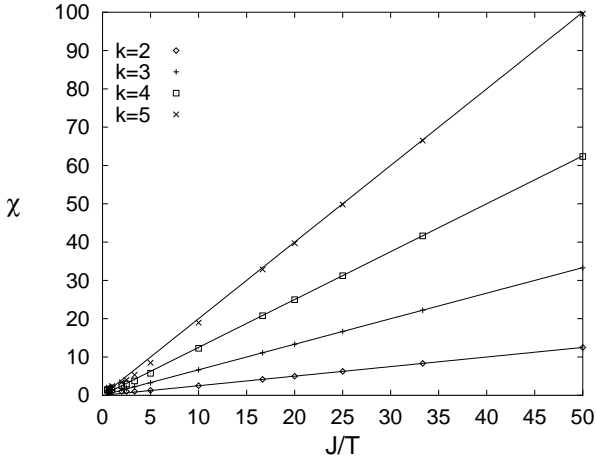}
\vspace{3mm}

\caption{Computed magnetic susceptibility for $k=2,\ldots,5$ (symbols)
and the paramagnetic contribution (\ref{chip_def}) (lines)}
\label{fig_chivstr}
\end{figure}

There are two contributions to the magnetic susceptibility:
\begin{enumerate}
\item The paramagnetic contribution, which is
\begin{equation}
\label{chip_def}
\chi_p=\frac{1}{3T}s_p(s_p+1)
\end{equation}
for low temperatures.
\item The contribution $\chi_a$ due to the antiferromagnetic correlations
inside the
chain fragment. Though the exact form of this contribution is unknown, it is
expected to be approximately linear in $k$ and only slowly varying as a
function of temperature.
\end{enumerate}
Figure \ref{fig_chivstr} shows the magnetic susceptibility as a function of
inverse temperature for fragment lengths $k=2,\ldots,5$. The numerically
computed values (various symbols) are almost perfectly connected by the
exactly known paramagnetic contribution (\ref{chip_def}) (lines), where
$(k-1)/2$ has been inserted for $s_p$. Please note that there are no fit
parameters.
\begin{figure}
\epsfxsize=85mm
\epsffile{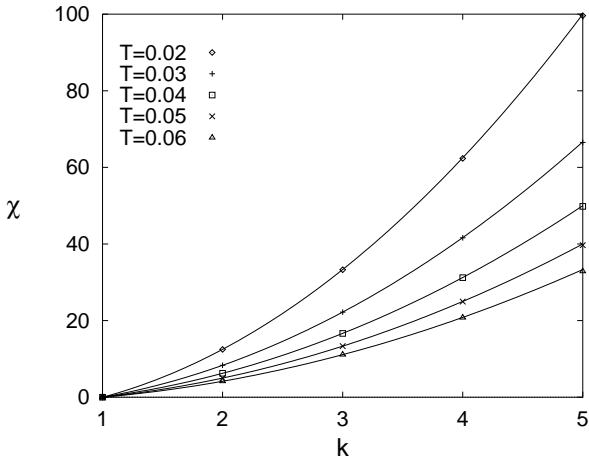}
\vspace{3mm}

\caption{Computed magnetic susceptibility for $T=0.02,\ldots,0.06$ (symbols)
and the paramagnetic contribution (\ref{chip_def}) (lines)}
\label{fig_chivskl}
\end{figure}
Therefore, at least for low temperatures, the susceptibility is well
described by
the Curie law (\ref{chip_def}). For higher temperatures $\chi_p$ is not
strictly linear in $T^{-1}$. Additionally, with increasing temperature
the antiferromagnetic contribution $\chi_a$ decays slower than $\chi_p$,
so $\chi_a$ may become relevant for larger $T$.

Of course, the coincidence of $\chi$ with $\chi_p$ for low temperatures
can also be
observed when plotting $\chi$ as a function of the fragment length $k$.
This is shown in figure \ref{fig_chivskl} for temperatures
$0.02\leq T \leq 0.06$.

\begin{figure}
\epsfxsize=85mm
\epsffile{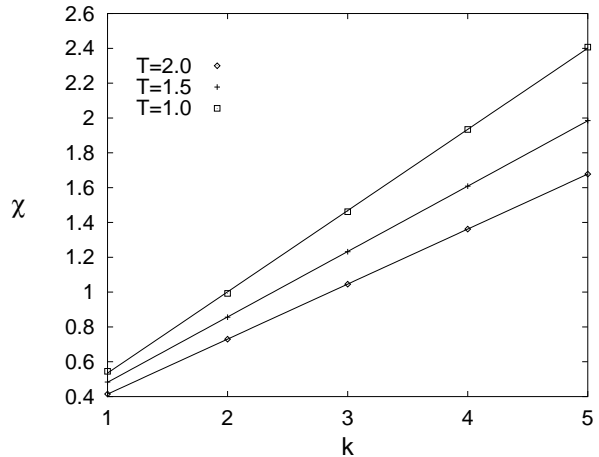}
\vspace{3mm}

\caption{Computed magnetic susceptibility for $T=1.0,1.5,2.0$ (symbols)
and fitted affine functions (\ref{chih_def}) (lines)}
\label{fig_chivskh}
\end{figure}

For high temperatures, the correlation length becomes smaller than the
fragment length. In this case, the effective total spin can be interpreted
as being composed of several independent blocks of individual spins.
The typical length of these blocks is the correlation length $\xi$, so
the effective total spin square $s_p(s_p+1)$
is proportional to $\xi k$, in contrast to the low temperature situation,
where $s_p=(k-1)/2$. This results in a {\em linear} $\chi$ vs $k$ dependence,
which is numerically confirmed, as shown in 
figure \ref{fig_chivskh}. There we have successfully fitted affine functions
\begin{equation}
\label{chih_def}
\chi_h=a_0(T) + a_1(T)\, k
\end{equation}
to the computed data points.
If we would extend figure \ref{fig_chivskl} to large values of $k$, we
would certainly observe a crossover from parabolic to linear $k$-dependence
at sufficiently large fragment length $k$. The crossover region depends on
temperature.

\begin{figure}
\epsfxsize=85mm
\epsffile{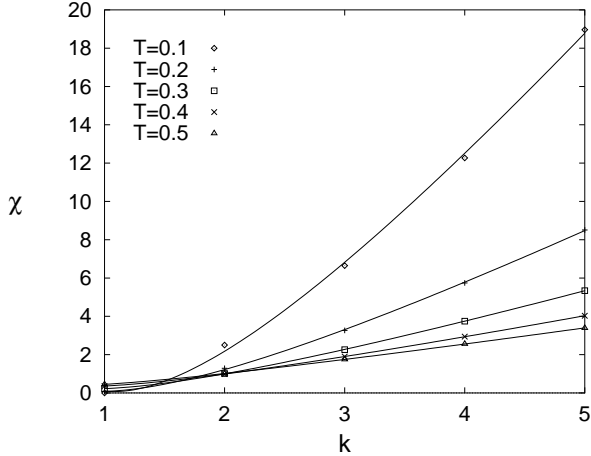}
\vspace{3mm}

\caption{Computed magnetic susceptibility for intermediate temperatures
(symbols) and fitted functions (\ref{chii_def}) (lines)}
\label{fig_chivski}
\end{figure}

We finish this section by considering intermediate temperatures, where the
crossover behaviour can be observed at finite fragment length $k<6$.
Figure \ref{fig_chivski} shows this intermediate temperature region. The
crossover from quadratic to linear behaviour is governed by a logarithmic
correction
\begin{equation}
\label{chii_def}
\chi_i=b_0(T_i) + b_1(T_i)\, k + b_2(T_i)\,\ln k,
\end{equation}
which fits $\chi(k)$ successfully for temperatures around $T=0.3$.
This behaviour presumably is reminiscent of the crossover in the free energy,
which takes place at the same temperature range.
\subsection{Mixed chains as ensembles of chain fragments}
This part of the paper is some kind of a synthesis of the analytic approach
developed in Section III and numerical computations performed in Section IV.
We shall illustrate this by computing the specific heat for
two temperatures, low and intermediate, $T=0.02$ and $T=0.3$, and for
reasons, which will be explained below, the
magnetic susceptibility versus ${\cal J}_0$ will be calculated for
intermediate temperatures $T=0.1$ and $T=0.3$.
Please note that in Section II we derived an analytical expression for the
magnetic susceptibility, which holds in the vicinity of the
$\langle 0\rangle\rightarrow\langle 1\rangle$ transition, where all
contributions of chain fragments of length $k\geq 2$ can be neglected,
i.e.\ at low enough temperatures.

The specific heat per compound spin is calculated as
\begin{eqnarray}
C=-T\frac{\partial^2 f}{\partial T^2},
\nonumber
\end{eqnarray}
where $f$ is the total free energy of the alternating chain.
In the thermodynamic limit, the partition function is given by the
logarithm of the maximum root $\lambda_{\rm max}$ of the polynomial
(\ref{eigenvalue}). The ``reference state'' in the derivation of the
partition function is the perfect $\langle\infty\rangle$-structure, so
the complete free energy is given by
\begin{equation}
f=-T\;\ln\,(\lambda_{\rm max}) + f_{\infty} .
\end{equation}
According to Table~I we have the estimate
\[ |f_{\rm int}(k)| \ll 0.02 \mbox{ for } k\geq 3. \]
Therefore it is sufficient to work with four different Boltzmann weights,
$w_0,\ldots,w_3$. (At $T=0.3$, this approximation is even more accurate.)
The polynomial (\ref{eigenvalue}) specializes to
\begin{eqnarray}
\lambda^4\! - (1\!+\!2w_0) \lambda^3\! + (2w_0\!-w_1) \lambda^2 \!+ 
(w_1-w_2) \lambda \nonumber\\+ (w_2\!-w_3)=0.\nonumber
\nonumber
\end{eqnarray}
Values for $f_{\infty}$ and $w_0,\ldots,w_3$ have been obtained from the
numerical results outlined in subsection A.
\begin{figure}
\epsfxsize=85mm
\epsffile{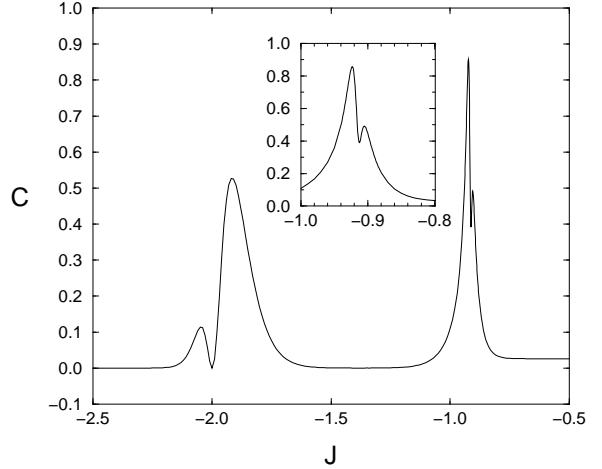}
\caption{Specific heat of the full alternating chain as a function of
${\cal J}_0$ at $T=0.02$. The insert shows the fine double peak structure
around ${\cal J}_0^{(1,\infty)}$. Note that for ${\cal J}_0>-0.7$ the specific
heat approaches a non-zero value (temperature dependent!), 
which is the specific heat of the perfect alternating chain 
$\dots-1/2-1-1/2-1-\dots$.}
\label{fig_fullclow}
\end{figure}
\begin{figure}
\epsfxsize=85mm
\epsffile{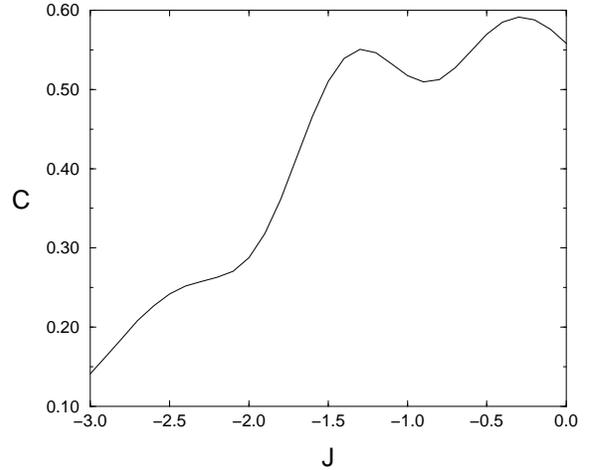}
\caption{Specific heat of the full alternating chain as a function of
${\cal J}_0$ at $T=0.3$. It decays to zero for large negative ${\cal J}_0$,
e.g. $C\approx 0.02$ at ${\cal J}_0=-4$.}
\label{fig_fullchigh}
\end{figure}

Figure \ref{fig_fullclow} shows the resulting specific heat as a function
of ${\cal J}_0$ for $T=0.02$. As expected, the most interesting features
can be found in the regions around ${\cal J}_0^{(0,1)}$ and
${\cal J}_0^{(1,\infty)}$. In order to understand the double-peak structure,
which is shown in the insert, consider the expression
\[
f(T)=\sum_{k\geq 0} x_k(T)\,\phi_k(T)
\]
for the total free energy of the alternating chain. As indicated, both the
concentrations $x_k$ and the free energies $\phi_k$ of the finite fragments
depend on temperature, so the specific heat involves $x_k$ itself and its
first and second derivatives. At low temperatures, $x_k$ is a step-like
function, which qualitatively explains the splittings of the peaks and 
their shapes.

At $T=0.3$, temperature has smeared out the fine structure around 
${\cal J}_0^{(0,1)}$ and ${\cal J}_0^{(1,\infty)}$. Both regions are not
isolated from each other, nevertheless the double-peak structure is still
visible in figure \ref{fig_fullchigh}.

Figures \ref{fig_fullclow} and \ref{fig_fullchigh} illustrate the dependence
of the specific heat on the interior coupling constant ${\cal J}_0$. The
temperature-dependence of $C$ will be published in future work.

Calculation of the magnetic susceptibility per compound spin is
straightforward. It is simply given by
\begin{equation}
\label{full_chi_sum}
\chi = \sum_{k\geq 0} x_k\,\chi_k,
\end{equation}
where $\chi_k$ is the susceptibility of a finite chain fragment of length $k$.
The concentrations $x_k$ according to (\ref{bm_c}) and (\ref{bm_more_k})
are computed by specializing to only four different Boltzmann weights.
$\chi_0$ is the susceptibility of an isolated paramagnetic spin-1/2. In units
of $\mu_B^2$, $\chi_0 = 1/4T$.
$\chi_1,\ldots,\chi_5$ are known numerically from subsection~A.
For ${\cal J}_0 < {\cal J}_0^{(1,\infty)}$ the largest eigenvalue
$\lambda_{\rm max} \gg 1$, so because of (\ref{bm_more_k}) we can safely
cut-off the summation (\ref{full_chi_sum}) after $k=5$. As
${\cal J}_0$ approaches ${\cal J}_0^{(1,\infty)}$, all concentrations $x_k$
for finite $k$ quickly decay to zero, as $\langle\infty\rangle$ becomes the
dominating configuration. The susceptibility $\chi_{\infty}$ of this
configuration is unknown for $T=0.02$, but we can extrapolate it easily
for those temperatures at which $\chi$ versus $k$ is in the linear regime.
From the calculations performed in subsection B, $\chi_{\infty}$ per compound
spin can be estimated as
\begin{eqnarray}
\begin{array}{l}
T=0.1:\quad\chi'=\lim_{k\to\infty} \frac{1}{k} \chi_k \approx 6.70,\\
T=0.3:\quad\chi'=\lim_{k\to\infty} \frac{1}{k} \chi_k \approx 1.60.
\end{array}
\label{extra_x_0.3}
\end{eqnarray}
Unfortunately, the convergence of the series (\ref{full_chi_sum}) is very bad
if ${\cal J}_0>{\cal J}_0^{(1,\infty)}$. At intermediate temperatures, this
problem can be circumvented by making use of the sum rule (\ref{sr}) and the
relations
\[\begin{array}{ll}
\lambda_{\rm max} \to 1 & \mbox{for increasing } {\cal J}_0 , \\
x_k \propto \lambda_{\rm max}^{-k} & \mbox{for large } k \mbox{ and } \\
\chi_k \propto k & \mbox{for large } k.
\end{array}\]
This procedure is illustrated in Appendix B.
However, in order to obtain reasonable results one needs to know the whole set
of $\chi_k$ before the $k$-dependence becomes linear.
There is no problem in case of temperatures $T=0.3$ and
$T=0.1$ (cf. figure \ref{fig_chivski}), but the set of $\chi_k$ at $T=0.02$
is too far from the linear regime (see figure \ref{fig_chivskl}).

\begin{figure}
\epsfxsize=85mm
\epsffile{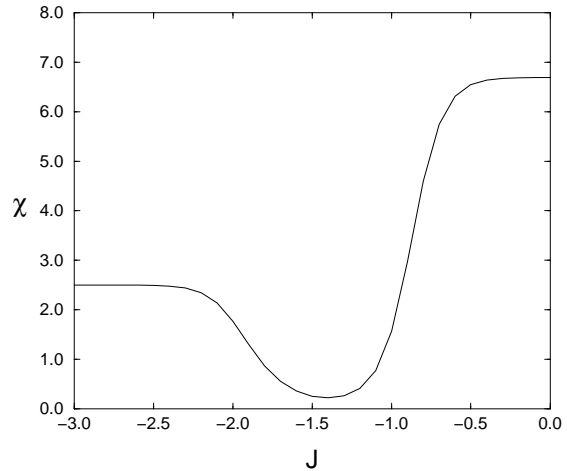}
\caption{Magnetic susceptibility of the full alternating chain as a function of
${\cal J}_0$ at $T=0.1$.}
\label{fig_fullxlow}
\end{figure}

\begin{figure}
\epsfxsize=85mm
\epsffile{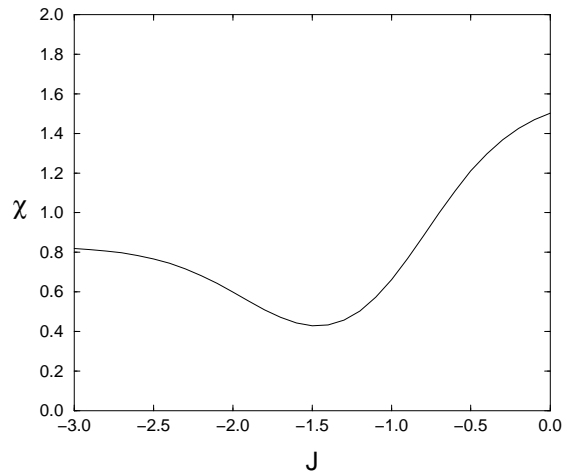}
\caption{Magnetic susceptibility of the full alternating chain as a function of
${\cal J}_0$ at $T=0.3$.}
\label{fig_fullxhigh}
\end{figure}

In figures \ref{fig_fullxlow} and \ref{fig_fullxhigh}, 
which show the magnetic susceptibility for
$T=0.1$ and $T=0.3$, we have used figure \ref{fig_chivski} as a prerequisite,
i.e. we extracted the fragment susceptibilities $\chi_0,\ldots,\chi_5$, and
the asymptotic slope $\chi'$.
Figures \ref{fig_fullxlow} and \ref{fig_fullxhigh} have the correct
asymptotics: At large negative ${\cal J}_0$ the susceptibility approaches
$\frac{1}{4T}$, at large positive ${\cal J}_0$ it is in accordance with
(\ref{extra_x_0.3}).

\section{Summary and outlook}
In this work we have combined analytical and numerical methods for calculating
thermodynamical properties of alternating Heisenberg chains. 
Interest in these spin systems is not purely academic. In fact, recent
progress in the observation of quantum effects in quasi one-dimensional
antiferromagnets, see e.g.\ \cite{br} and \cite{gntsb},
is commonly contributed by physicists and crystal growth experts. It
is out of our goal to recommend some definite compounds, which might
be good candidates for mixed chains with such a competitive
behaviour. Schematically, their simplest representation is given in
our description of models A and B (see also illustrations in \cite{nzu}).

It is necessary to emphasize, that the unusual thermodynamic behaviour
described in this paper can be efficiently observed in those
compounds, where the values of the coupling constants $J_0$ and $J_1$ 
lead to competitive interactions. Then a major contribution to
thermodynamic quantities comes from thermal changes of the fragment
concentrations $x_k$.
\begin{enumerate}
\item 
To make the basic ideas transparent, we employ a simple representation
of these chains: ``compound spins'' and spin-1/2 sites alternate
within the chain. They interact via a coupling constant ${\cal J}_1$.
A compound spin consists of two spin-1/2 sites forming a dumbbell
configuration perpendicular to the chain direction. 
The constituents of a dumbbell interact via the internal coupling
constant ${\cal J}_0$.
The intrinsic state of a compound spin is either spin-1 or spin-0,
which gives rise to the hidden Ising symmetry of the Heisenberg
chain. Spin-0 states can be treated as
``defects'' with respect to the perfect alternating chain
$\ldots -1/2-1-1/2-1-\ldots$.
These equilibrium ``defects'' simply break the perfect chain into a set of 
non-interacting chain fragments of finite lengths, whose general formula
is $(\frac{1}{2},1)^k \frac{1}{2}$ for length $k\geq 0$.
\item
At zero temperature, one of three ground states is realized, depending
on the value of ${\cal J}_0$ (${\cal J}_1$, the antiferromagnetic exchange,
is set to $-1$): \\
$\langle 0 \rangle$, the regular structure of chain fragments of length 
$k\!=\!0$ for ${\cal J}_0\!<\!-2$, this means that
all compound spins are zero; \\
$\langle 1 \rangle$, the regular structure of chain fragments of length 
$k\!=\!1$ for $-2\!<\!{\cal J}_0\!<\!-0.910$, i.e.\ compound spins 0 and 1
alternate with each other; \\
$\langle \infty \rangle$, only one ``fragment'' of infinite length
for $-0.910\!<\!{\cal J}_0$, i.e.\ all compound spins are 1. \\
Although the first order transitions governed by ${\cal J}_0$ 
are smeared out at non-zero 
temperature, thermodynamic quantities display a strong dependence on
${\cal J}_0$ and temperature in the vicinities of the zero-temperature
critical values.
\item As a function of ${\cal J}_0$, the specific heat displays two peaks,
which are reminiscent of the zero-temperature transitions
$\langle 0 \rangle\to\langle 1 \rangle\to\langle\infty \rangle$.
As shown in figure \ref{fig_fullclow}, these peaks are split at low
temperatures. The origin of this fine structure is the step-like
behaviour of the fragment concentrations $x_k$, predominantly
$x_0$ and $x_1$.
With increasing temperature, the peaks become broader and overlap,
but the local minima still point out the vicinities of the critical
${\cal J}_0$ values, cf. figure \ref{fig_fullchigh}.
The crucial dependence of the magnetic susceptibility 
on ${\cal J}_0$ and $T$ is illustrated in figures \ref{fig_fullxlow} and
\ref{fig_fullxhigh}. Technically it is more difficult to compute 
$\chi({\cal J}_0)$ because of the poor convergence of series 
(\ref{full_chi_sum}). We have eliminated these difficulties by making
use of the sum rule (\ref{sr}).
\item All these pretransitional phenomena, which can be identified clearly,
are provided by the configurational contributions to thermodynamic quantities.
We have developed an analytic approach for calculating the free energy and
the concentrations of chain fragments of different lengths.
This is achieved by solving polynomial equations,
whose degree increases at lower temperature.
In fact, the hierarchical structure of the equations is
regulated by $f_{\rm int}(k)$, i.e.\ the part of the free energy
of a chain fragment of length $k$ which can be interpreted as the
interaction of fragment ends. From numerical computations we observed that
$f_{\rm int}$ decays rapidly with $k$.
\item For other alternating chains, where the non-compound sites are occupied
by higher spins $s$ instead of $1/2$, there are several succesive
zero-temperature transitions with increasing ${\cal J}_0$.
The last one is $\langle 4\rangle \to \langle \infty\rangle$ if $s\geq 3/2$.
For these chains, the occurence of several peaks in the
specific heat would be possible. However, very low temperatures are needed
to isolate some peaks: Although the successive transitions
$\langle 0\rangle \to \langle 1\rangle$ and 
$\langle 1\rangle \to \langle 2\rangle$ are sufficiently separated,
the distance between $\langle 3\rangle \to \langle 4\rangle$ and
$\langle 4\rangle \to \langle \infty\rangle$ is only
$\Delta{\cal J}_0\sim 10^{-3}$.
\item To investigate the thermodynamics of alternating chains with
$s\!>\!1/2$, other methods could be used instead of exact numerical
diagonalization. A modification of the spin-wave approximation, which
can be adjusted to our problem, has been proposed by M. Takahashi \cite{taka}.
In \cite{klue1} and \cite{klue2}, the quantum transfer matrix of exactly
solvable spin chains has been investigated in order to calculate thermodynamic
quantities. A combination of this approach and numerical methods could be
applied to our problem.
\item Generalization to higher dimensions is straightforward: For instance
in two dimensions, the ``white'' squares of a checkerboard lattice contain
single spins, while the ``black'' ones are occupied by compound
spins. Competition may again reveal a hidden Ising-like variable and
possible two-dimensional superstructures. Of course, similar models can be
constructed on any bipartite lattice.
\end{enumerate}
\subsection*{Acknowledgement}

This work has been performed within the research program of the
Sonderforschungsbereich 341, K\"oln-Aachen-J\"ulich. 
\section*{Appendix A}
\vspace{-2mm}

\renewcommand{\theequation}{A.\arabic{equation}}
\setcounter{equation}{0}
As shown in section IV, $f_{\rm int}(k)$ decays rapidly at high temperatures. 
But even at very low temperatures, $T=0.02$, we have the estimate
\[ |f_{\rm int}(k)|\ll T \quad{\rm for}\quad k\geq 3. \]
According to the method developed in section III, it is
sufficient to introduce the set of Boltzmann weights:
\begin{eqnarray}
\{w_1,\, w_2=w_1\exp[(f_{\rm int}(1)-f_{\rm int}(2))/T],\nonumber\\ 
w_3=w_1\exp[f_{\rm int}(1)/T],\, w_n=w_3 (n\geq 3)\}.\nonumber
\end{eqnarray}

In this appendix we outline the procedure for the general case, where all
statistical weights (\ref{set_w}) are taken into account.

If we substitute $n$ by $n-1$ in (\ref{z3}) and subtract the resulting
expression from the original equation (\ref{z3}) we arrive at
\begin{eqnarray}
{\cal Z}_{n+1}\! - (1\!+2w_0){\cal Z}_n\! +(2w_0\! - w_1) {\cal Z}_{n-1} + 
(w_1\! - w_2){\cal Z}_{n-2}
\nonumber\\
+ (w_2\! - w_3){\cal Z}_{n-3} +\cdots +(w_{k-1}\! - w_k){\cal Z}_{n-k}=0,\nonumber
\end{eqnarray}
which is a generalization of (\ref{z4}). It can be rewritten in compact
form
\begin{eqnarray}
{\cal R}_{k+1}{\cal Z}_{n+1}=0,\;n\geq k, 
\label{z_app}
\end{eqnarray}
if the linear operator
\begin{eqnarray}
{\cal R}_{k+1}f_{n+1}\equiv f_{n+1}\!-(1\!+\!2w_0)f_{n}\!+(2w_0\! -w_1)f_{n-1}\nonumber\\
+(w_{1}-w_{2})f_{n-2} +\cdots +
(w_{k-1}-w_{k})f_{n-k}
\label{r}
\end{eqnarray}
is introduced.
Directly related to the operator ${\cal R}_{k+1}$ is the characteristic
equation
\begin{eqnarray}
P_{k+1}(\lambda)\equiv\lambda^{k+1}-(1+2w_0)\lambda^{k}+(2w_0-w_1)\lambda^{k-1}\nonumber\\
+(w_{1}-w_{2})\lambda^{k-2} +\cdots +
(w_{k-1}-w_{k})=0
\label{eigenvalue}
\end{eqnarray}
whose maximal root, $\lambda_{\rm max}$, from the set $\{\lambda_1,\dots,
\lambda_{k+1}\}$
determines the asymptotic behaviour of the partition function.

Analogous equations for the ${\cal D}^{(m)}$ functions can be derived
and written in compact form as
\begin{eqnarray}
{\cal R}_{k+1}{\cal D}_{n+1}^{(m)}=w_m({\cal Z}_{n-m}-{\cal Z}_{n-m-1}),
\label{d_app} 
\end{eqnarray}
where $w_m$ must be substituted by $2w_0$ for $m\!=\!0$.

In our formalism it is sufficient to find out the 
${\cal D}^{(m)}$ functions up to rank $k$ ($m=0,1,\dots,k$).
For higher rank $m\!>\!k$ one obtains
\begin{eqnarray}
\label{d_k_shift}
{\cal D}_{n+m-k}^{(m)}={\cal D}_n^{(k)}.
\end{eqnarray}
This relationship becomes clear from the form of 
Eqs.~(\ref{d_app}) for $m\geq k$, where $w_m=w_k$.

An obvious ansatz to solve Eqs.~(\ref{z_app}) is
\begin{eqnarray}
{\cal Z}_n=\sum_{i=1}^{k+1}c_i\lambda_i^n,
\label{z_sol}
\end{eqnarray}
whereas for solutions of Eqs.~(\ref{d_app}) we use
\begin{eqnarray}
{\cal D}_n^{(r)}=\sum_{i=1}^{k+1}(a_i^{(r)}+b_i^{(r)}n)\lambda_i^n,\quad
r=0,1,\dots,k.
\label{d_sol}
\end{eqnarray}
We do not need the complete set of coefficients $\{a_i^{(r)},b_i^{(r)},c_i\}$,
which can be calculated by using the boundary conditions, 
but rather the set of ratios $\{b_{\rm max}^{(m)}/c_{\rm max}\}$,
which yields the concentrations $x_m$. 

After inserting the ansatz (\ref{d_sol}) into Eq.~(\ref{d_app}),
we compare the leading terms, which are given in terms of powers of 
$\lambda_{\rm max}$. Neglecting all other terms is equivalent to taking
the thermodynamic limit. The basic result for the concentrations
$x_m$ reads
\begin{eqnarray}
\label{bm_c}
x_m=\frac{b_+^{(m)}}{c_+}=
\frac {w_m(\lambda_{\rm max}-1)\lambda_{\rm max}^{k-m-1}}
{Q_{k+1}(\lambda_{\rm max})},
\end{eqnarray}
where 
\begin{eqnarray}
Q_{k+1}(\lambda)\! =\! \lambda^{k+1}\! - \!(2w_0\!-w_1)\lambda^{k-1}\nonumber\\ 
- 2 (w_1\!-w_2)\lambda^{k-2} -\cdots -k(w_{k-1}-w_k)\nonumber\\
\equiv \lambda^{k+1} \frac {d}{d\lambda}
\left(\frac{P_{k+1}(\lambda)}{\lambda^k}\right).
\label{Q}
\end{eqnarray}
Combining Eq.~(\ref{d_k_shift}) with result (\ref{bm_c}) yields
\begin{eqnarray}
\label{bm_more_k}
x_{k+m}=\frac 1{\lambda_{\rm max}^{m}}x_k.
\end{eqnarray}
\section*{Appendix B}
\vspace{-2mm}

\renewcommand{\theequation}{B.\arabic{equation}}
\setcounter{equation}{0}
For the temperatures of interest we can write the susceptibility
in the form
\begin{eqnarray}
\chi=x_0\,\chi_0 +x_1\,\chi_1 +x_2\,\chi_2 +x_3\,(\chi_3 +
\chi_4\,\lambda_{\rm max}^{-1}\nonumber\\ + 
\chi_5\,(\lambda_{\rm max}^{-2} +\lambda_{\rm max}^{-3}+\dots)
\nonumber\\
+\chi'\,\lambda_{\rm max}^{-3}(1+2\lambda_{\rm max}^{-1}+
3\lambda_{\rm max}^{-2}+\dots)).
\label{chi}
\end{eqnarray}
The most singular behaviour is contained in the term which is proportional
to $\chi'$.
It can be accurately evaluated by using the sum rule (\ref{sr}). Let us
reorganize the terms of (\ref{chi}) into three groups. The first one
contains the non-singular contributions
\begin{eqnarray}
x_0\,\chi_0 +x_1\,\chi_1 +x_2\,\chi_2 +
x_3\,(\chi_3 +\chi_4\,\lambda_{\rm max}^{-1}\nonumber\\
+\chi'\,\lambda_{\rm max}^{-3}(1+2\lambda_{\rm max}^{-1}+
3\lambda_{\rm max}^{-2})).
\label{chi_1}
\end{eqnarray}
The second one is pseudo-singular:
\begin{eqnarray}
\frac{x_3\,\chi_5}{\lambda_{\rm max}(\lambda_{\rm max}-1)}.
\label{chi_2}
\end{eqnarray}
Although the denominator in (\ref{chi_2}) vanishes if $\lambda_{\rm max}\to 1$,
the complete term remains finite, since $x_3\to 0$ in this limit, too.
(cf.\ Eq.\ (\ref{bm_c}).) Due to the sum rule
\begin{eqnarray}
x_0 + 2x_1 + 3x_2 +x_3\,(4+5\lambda_{\rm max}^{-1}+
6\lambda_{\rm max}^{-2}+\dots)=1 ,
\nonumber
\end{eqnarray}
the last contribution
\begin{eqnarray}
x_3\,\chi'\,\lambda_{\rm max}^{-6}(4+5\lambda_{\rm max}^{-1}+
6\lambda_{\rm max}^{-2}+\dots)
\nonumber
\end{eqnarray}
can be rewritten in the form
\begin{eqnarray}
\chi'\,\lambda_{\rm max}^{-6}(1-x_0 -2x_1 -3x_2),
\label{chi_3}
\end{eqnarray}
which evidently approaches $\chi'$ when $\lambda_{\rm max}\to 1$.

\end{multicols}
\end{document}